\documentclass[prl,floatfix,twocolumn]{revtex4}
\usepackage{amsmath}
\usepackage{amsfonts}
\usepackage{mathrsfs}
\usepackage{textcomp}
\usepackage{graphicx}
\usepackage{epstopdf}
\usepackage{hyperref}
\usepackage{color}
\usepackage{dcolumn}
\usepackage{commath}
\usepackage{url}
\usepackage{gensymb}
\usepackage{siunitx}
\usepackage{doi}

\begin{document}

\title{Frequency gradients in heterogeneous oscillatory media can 
spatially localize self-organized wave sources that coordinate
system-wide activity}
\author{Ria~Ghosh$^{1,2}$, Pavithraa~Seenivasan$^{1}$, Shakti~N.~Menon$^{1}$, S.~Sridhar$^{1,3}$, 
Nicolas~B.~Garnier$^{3}$, Alain~Pumir$^{3}$, and Sitabhra~Sinha$^{1,2}$}
\affiliation{$^{1}$The Institute of Mathematical Sciences, CIT Campus, Taramani, Chennai 600113, India}
\affiliation{$^{2}$Homi Bhabha National Institute, Anushaktinagar, Mumbai 400094, India}
\affiliation{$^{3}$Robert Bosch Centre for Cyber-Physical Systems, Indian Institute of Science, Bengaluru 560012, India}
\affiliation{$^{4}$Universit\'{e} de Lyon, ENS de Lyon, Universit\'{e}
Claude Bernard, CNRS, Laboratoire de Physique, F-69342 Lyon, France}

\begin{abstract}
Rhythmogenesis, which is critical for many biological functions, involves a transition to coherent activity through cell-cell communication. 
In the absence of centralized coordination by specialized cells 
(pacemakers), competing oscillating clusters impede this global synchrony. 
We show that spatial symmetry-breaking through a frequency gradient 
results in the emergence of localized wave sources driving system-wide 
activity. Such gradients, arising through heterogeneous inter-cellular 
coupling, may explain directed rhythmic activity during labor in the 
uterus despite the absence of pacemakers.
\end{abstract}

\maketitle

\begin{figure}[tbp]
\begin{center}
\includegraphics[width=0.99\columnwidth]{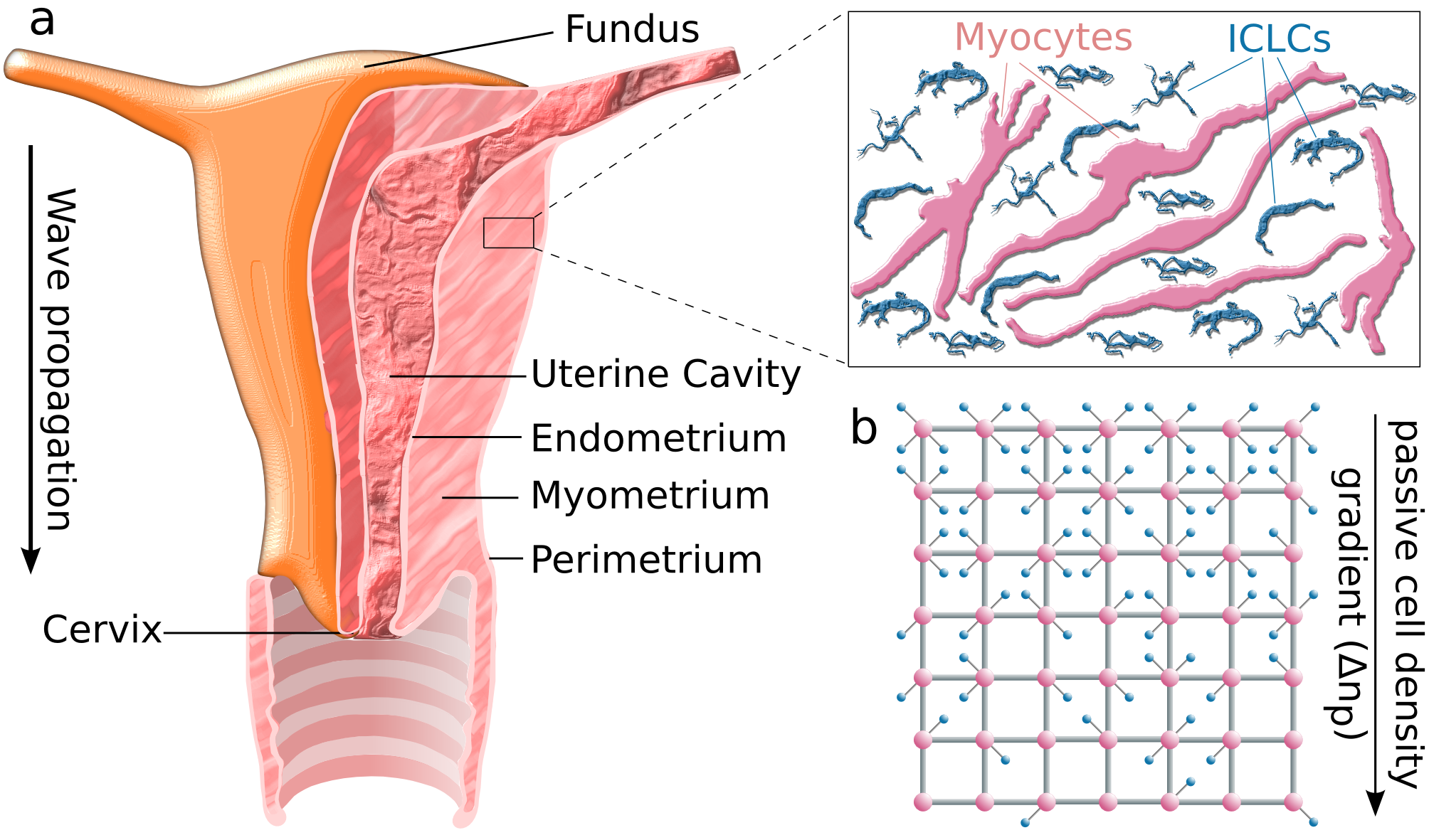}
\end{center}
\caption{(a) Schematic diagram of a human uterus, which immediately before
parturition exhibits coherent excitation activity appearing to
propagate from the fundus to the cervix.
A cross-section along
the uterine wall is shown to indicate the myometrium, whose
heterogeneous cellular composition, comprising excitable (myocytes)
and passive (ICLCs) cells is shown in the inset
[adapted from Refs.~\citep{Yoshino1997} and \citep{Duquette2005},
respectively, and drawn to scale]. 
(b) The uterine myometrium is
modeled as a two-dimensional square
lattice with each site occupied by an excitable cell coupled to
passive cells, whose density is proposed to decrease monotonically
along the organ from the fundus to the cervix
(i.e., along the vertical axis of the lattice).}
\label{fig_schematics}
\end{figure}

Rhythmic activity in many natural systems is centrally coordinated
by specialized ``pacemakers''~\cite{Smith1991,Czeisler1999,Koshiya1999,Young2001,Reppert2002,Lincoln2006,Rabinovich2006,Morquette2015}, examples ranging
from the sino-atrial node of the 
heart~\cite{DiFrancesco1986,DiFrancesco1993} to the interstitial cells of Cajal
in the gastro-intestinal tract~\cite{Huizinga1995,Thomson1998,Huizinga2014}.
However, functionally critical rhythmic contractions can also appear in
organs such as the uterus where
no pacemaker cells have been identified~\cite{Smith2015}. In such cases, 
self-organized synchronization of activity can arise through cells communicating
with their neighbors~\cite{Cartwright2000,Boschi2001,Singh2012,Xu2015}. However, this could lead to
multiple oscillating clusters characterized by distinct frequencies and 
phases to co-exist in different locations in the medium.
The potential conflict between these competing coordination centers can
prevent coherence~\cite{Lee1996}.
In normal labor, regular contractions of the uterus progressively dilate
the cervix, eventually resulting in natural delivery. This
requires rhythmic wave-like activity that propagates along
the myometrium to be directed from a source located
near the fundus [Fig.~\ref{fig_schematics}~(a)]~\cite{CaldeyroBarcia1950,Reynolds1951,Wolfs1970,Planes1984,
Buhimschi2009,Euliano2009}. Understanding
how such apparent ``fundal dominance''~\cite{Alvarez1954} comes about, despite the absence
of any specialized group of pacemaker cells in the uterus, is important
as labor dystocia, involving abnormally slow progress of labor or 
its complete arrest, is a significant cause of maternal as well as 
fetal morbidity~\cite{Neilson2003}. Currently, the treatment of 
such birth disorders involve non-elective primary 
cesarean delivery 
(accounting for more than a third of such procedures in USA~\cite{Barber2011}),
which increases the risk of adverse maternal and neonatal 
outcomes~\cite{Clark2011,Caughey2014}.
Elucidating the mechanisms
by which coordination over the entire organ is achieved prior to 
parturition can help in
devising safer intervention methods.

From the perspective of dynamical systems, the problem is one of breaking
the spatial symmetry so that certain
regions impose their rhythmic pattern on the
rest of the system by means of excitation wavefronts that initiate contraction
as they propagate through the medium. Thus, although arising through 
self-organization, such a system may appear to possess a ``pacemaker'' 
region from which waves activating the rest of the medium originate.
A feasible symmetry-breaking mechanism in biological oscillatory medium
is to have a spatial gradient in the frequency
of periodic activity. Indeed, we note that such
a frequency gradient is known to exist for slow waves in the small
intestine which provides polarity to peristaltic contractions~\cite{Alvarez1914,Bortoff1976,Ermentrout1984}. In the case of the
uterus, which has a highly heterogeneous cellular composition, the frequency 
of activity can be modulated by the 
coupling between excitable smooth muscle cells (myocytes) and electrically 
passive cells, such as fibroblasts and Interstitial Cajal-like cells (ICLC) 
[inset of Fig.~\ref{fig_schematics}~(a)]. 
Thus, a frequency gradient could arise from a 
variation in the density of
passive cells coupled to myocytes and/or the expression
of connexin proteins forming gap junctions that regulate the inter-cellular
coupling. This in turn could be the result of signaling molecules
diffusing from a source, which, if located
at the upper end of the uterus, e.g., near the fundus, will lead to
a gradually decreasing concentration of coupled passive cells and/or 
gap junctions. In this paper we investigate the consequences of the
existence of such a spatial symmetry-breaking gradient on the 
dynamics of the system. Specifically,
we show that a sufficiently steep gradient will result in the emergence
of one (or few) organizing centers of activity that are spatially 
localized in the medium. 
Thus, even in the absence of pacemaker cells, contractions will appear to
be coordinated by waves emanating from a source whose location, depending
on the gradient, can be near the fundus.

To model dynamical activity in gravid uterine tissue, we note that
the number of gap junctions increases significantly over the course of 
pregnancy~\cite{Miyoshi1996,Garfield1981}, eventually promoting coordination
of periodic activity across the entire organ~\cite{Ramon2005,Garfield2007,Rabotti2015}. 
Following Refs.~\cite{Jacquemet2006,Singh2012}, we consider 
rhythmogenesis in 
this system to be a self-organized outcome of interacting heterogeneous
non-oscillating cells.
Indeed, in the uterus, no experimental evidence for specialized
pacemaker cells~\cite{Wray2001,Ramon2005} or for myocytes capable of auto-rhythmicity~\cite{Shmygol2007} has been found so far.
We describe the activity of a uterine myocyte in terms of the cellular 
transmembrane potential $V_e$ and effective conductance $g$ using the 
FitzHugh-Nagumo model~\cite{Keener1998}. This is a generic representation 
of the dynamics of an excitable system~\cite{Sinha2014}, which
qualitatively reproduces the behavior seen in more physiologically
realistic models of uterine activity~\cite{Xu2015}. The pair of coupled
equations specifying this model are:
$\dot{V_e}= F_e (V_e, g) = AV_e(V_e -\alpha)(1-V_e)-g$,
$\dot{g}= G (V_e, g) = \epsilon(V_e -g)$, where the parameters 
$A(=3), \alpha(=0.2)$ and $\epsilon(=0.08)$
govern the fast activation kinetics, excitation threshold and recovery rate,
respectively. The dynamical state of a passive cell is characterized by its membrane potential $V_p$ which evolves as $\dot{V_p}=F_p(V_p) = K(V_p^R -V_p)$ where $V_p^R (=1.5)$ is the resting state value of $V_p$ and $K (=0.25)$ is the corresponding timescale 
~\cite{Kohl1994}. 
We investigate the propagation of electrical activity across the organ
by considering a system comprising cells arranged in a
$2$-dimensional lattice of size $L \times L$. 
As the simulation domain is meant to represent electrical activity over the entire
uterine myometrium, we implement periodic boundary condition along the vertical edges
and no-flux along the horizontal edges of the lattice.
Each lattice point $i$ comprises an excitable
cell coupled to $n_p^i$ passive cells 
($n_p^i = 0,1,2...$) with strength $C_r$, while
neighboring excitable cells interact diffusively with
strength $D$. Thus,
the dynamics of the spatially extended system is described as:
\begin{align*}
d V_{e}^i/dt &= F_e(V_{e}^i, g^i) + n_p^i C_r (V_p^i - V_e^i)
+ D \sum_{\langle i,j \rangle} (V_e^j - V_e^i), \\
d g^i/dt &= G(V_{e}^i, g^i),\\
d V_{p}^i/dt &= K(V_{p}^{R} -V_{p}^i)-C_{r}(V_{p}^i -V_{e}^i),
\label{eq:final_form}
\end{align*}
where $\langle i,j \rangle$ represents the set of excitable cells that
neighbor $i$ ($i=1, \ldots L^2$).
For the simulations reported here we have chosen $L=450$. We have verified that our results are qualitatively similar for smaller lattice
sizes $L \geq 128$.
We assume
$D=C_r$ suggesting that gap-junctions are equally likely to couple
excitable cells with other excitable cells and to passive cells.
Note that, all the passive cells connected to a given excitable cell at $i$
behave identically and hence are represented using the single variable
$V_p^i$. The value of $n_p$ at each lattice site $(l,m)$ is randomly sampled from a Poisson distribution with parameter $\lambda (l,m)$, whose global average
is the ratio $f$ of the total number of passive cells to the
number of excitable cells in the lattice. 

Coupling excitable cells to passive ones can induce
autonomous periodic activity which is absent in isolation in 
either of the cell types~\cite{Jacquemet2006,Singh2012,Xu2013}.
The frequency of oscillations at each site $(l,m)$ depends in
a non-monotonic manner on $n_p$ (see Fig.~1~b of Ref.~\cite{Singh2012}), 
and
hence on $\lambda (l,m)$, as well as, 
the coupling constant $C_r (=D)$ (see Fig.~1~c of Ref.~\cite{Singh2012}).
Thus, by introducing a 
linear gradient in the passive cell density 
along the longitudinal axis of the domain, viz.,
$\lambda (l,m) = [l-(L/2)] \Delta n_p + n_p^{mid}$
[Fig.~\ref{fig_schematics}~(b)], we can obtain a systematic variation
in the frequency of oscillation across space. 
We have examined density gradients chosen from a range in which
the frequency will, on average, change monotonically 
across the longitudinal axis, viz., $0 \leq \Delta n_p \leq 9\times10^{-4}$. 
For a system size $L$ and slope $\Delta n_p$ of the gradient, 
we choose $n_p^{mid}$, the value of $n_p$ at the center of the domain,
such that $f = 0.7$, which ensures oscillatory behavior. We have verified that the behavior reported here is robust over multiple realizations with random initial conditions and passive cell distributions. We note that similar frequency gradients can arise from variation
in the gap-junction coupling strength ($D, C_r$) as a result of spatial 
heterogeneity in the expression of connexin proteins.

To model the increase in inter-cellular coupling in the gravid uterus 
over time~\cite{Smith2015}, we have adiabatically increased $C_r (=D)$ over the 
entire simulation domain after starting
from random initial conditions for a sufficiently low value of the 
coupling. For a dynamical system having multiple attractors, such an 
approach may yield strikingly different behavior for the evolving system
compared to the states observed when the coupling strengths are
temporally invariant. 

\begin{figure}[tbp]
\begin{center}
\includegraphics[width=0.99\columnwidth]{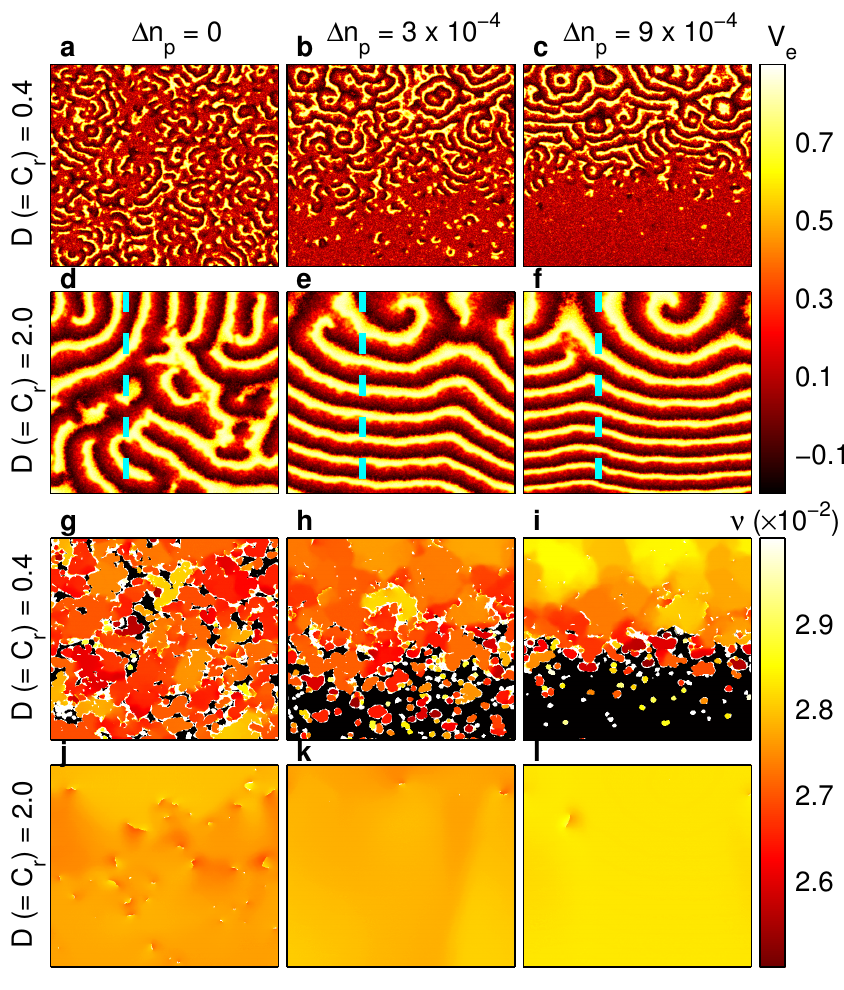}
\end{center}
\caption{Propagation of excitation waves along the medium gets more
organized with increasing slope $\Delta
n_p$ of the passive cell density gradient.
(a-f) Snapshots of the activity $V_e$
in a two-dimensional simulation domain ($L=450$) for two different
values of inter-cellular coupling strengths $D (=C_r)$ [first row:
$D=0.4$, second row: $D=2$]~\cite{note4}. The homogeneous system obtained in
the absence of the density gradient [left column: $\Delta
n_p=0$] is compared with the situations
seen for finite gradients [middle column: $\Delta n_p=3 \times
10^{-4}$ and right column:$\Delta n_p = 9 \times 10^{-4}$]. 
(g-l) The corresponding pseudocolor plots indicating the oscillation
frequencies of individual sites in the medium
(black: absence of oscillation).
For lower coupling strength [third row: $D=0.4$], increasing
passive cell density gradient results in
distinct frequency clusters merging with each other.
For sufficiently high gradient the medium is divided into a region
exhibiting activity (corresponding to higher passive cell density) and
an almost quiescent region (at lower density).
However, for higher coupling strength, viz., $D=2$ 
[fourth row: $\Delta n_p = 9 \times 10^{-4}$], the system exhibits
global synchronization with effectively a single frequency dominating
the activity in the entire medium. Localized phase defects correspond
to tips of spiral waves 
(second row). With increasing density gradient, we observe fewer spiral waves.
Furthermore, they are confined to the region having higher passive 
cell density, 
which appears as the source of
excitation fronts propagating across the domain.} 
\label{fig_annealed}
\end{figure}

In order to understand the effect of having a gradient in the density
of passive cells, and by extension, the frequency of local spontaneous
activation in the medium, we systematically vary the slope of the 
gradient $\Delta n_p$ and the inter-cellular coupling strength $D (= C_r)$.
Fig.~\ref{fig_annealed} shows snapshots of the collective dynamics,
represented by the field corresponding to the excitable cell transmembrane 
potential ($V_e$, panels a-f) and that of the local frequencies ($\nu$, panels g-l),
for different choices of $\Delta n_p$ and $D (= C_r)$.
In the absence of a gradient ($\Delta n_p=0$), the
activity is not confined to any particular spatial domain even at low
values of coupling (panel a). Through stochastic fluctuations certain
lattice sites have higher values of $n_p$ than their neighbors. As a result,
the oscillating activity in such a site dominates the local region on account of having the highest frequency ~\cite{Sinha2014}. 
Thus, we observe a large number of autonomous sources of excitation, 
each generating waves that are confined to its local region of influence.
Introducing a gradient with finite
$\Delta n_p$ (panels b and c)
results in a spatial heterogeneity
in the distribution of such excitation sources. At the upper end of the
gradient, sites will typically have higher frequencies (arising from
the larger values of $n_p$ on average) compared to the rest of the medium 
(panel h).
In contrast, $n_p$ is small at the lower end of the gradient, 
resulting
in either the absence of sources of oscillatory activity or ones
having relatively much lower frequencies. On increasing the gradient,
we observe the activity to become more localized (compare
panels b and c) and the spatial ordering of frequencies to become more
pronounced (see panels h and i).

\begin{figure}[tbp]
\begin{center}
\includegraphics[width=0.99\columnwidth]{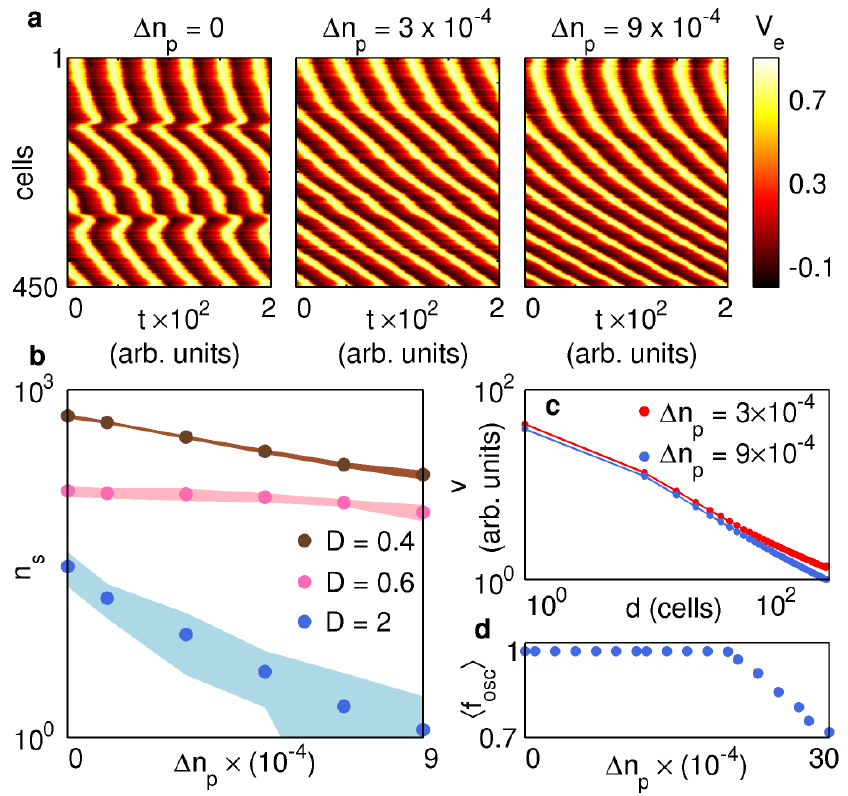}
\end{center}
\caption{Increasing passive cell density gradient promotes
unidirectional propagation of excitation waves through the medium.
(a) Spatio-temporal evolution of the activity $V_e$ for three different
values of the gradient $\Delta n_p$ in the medium 
along the broken line segments shown in
Fig.~\ref{fig_annealed}~(d-f) respectively. 
In the absence of a
gradient, the system exhibits multiple coordination centers,
characterized by excitation fronts propagating away from them.
Introducing a gradient in $n_p$ results in a sharp reduction in their 
number, with the waves emanating from centers that are localized
at the higher end of the gradient.
(b) Decrease in the number of phase singularities 
%corresponding to spiral wave tips 
with increasing $\Delta n_p$
shown for different values of inter-cellular
coupling $D (= C_r)$. 
The mean (filled circles) and standard deviation (shaded regions) of the
temporal average of the number of phase singularities $n_s$, 
calculated over an ensemble, are displayed.
(c) The phase velocity $v$ of the excitation front is measured along a 
one-dimensional chain with linear gradients in $n_p$, and 
decreases with distance $d$ from the higher end
of the gradient.
(d) The mean fraction of oscillating cells $f_{osc}$ in the medium obtained
by averaging over an ensemble shown
as a function of the slope $\Delta n_p$ of the passive cell density gradient.
}
\label{fig_kymo_annealed}
\end{figure}

When the inter-cellular coupling is increased (Fig.~\ref{fig_annealed}, d-f), 
we observe increased wavelength for the propagating excitation fronts and
in general, higher coordination in activity across regions.
This is apparent on comparing between the corresponding spatial 
distributions of frequencies in panels g-i (low coupling) and panels j-l
(high coupling). The clusters of distinct
frequencies seen for low $D (=C_r)$ (panel g) become homogenized upon
increasing the coupling (panel j), essentially resulting in
synchronization of oscillatory activity across the domain. 
We also note in this case (i.e., for $\Delta n_p=0$) the presence of rotating {\em spiral waves} which
are characterized by the existence of a phase singularity at the
spiral tip~\cite{Sinha2014}. 
The propagating fronts have no preferred orientation as the passive cell density is homogeneous on average.
There are also no constraints on where the spiral waves can occur,
which is
reflected in the coexistence of several competing organizing centers 
of wave activity. The introduction of a gradient in $n_p$ (panel e) breaks this
spatial symmetry, resulting in spiral waves at the upper end of the 
gradient activating the surrounding region at a frequency higher
than that elsewhere. The activity of
the entire domain is eventually enslaved to that of these organizing centers, 
which consequently are drastically reduced in number and spatially localized
at the fundal end. This is consistent with empirical observations of
multiple foci of activity in this region in different species~\cite{Lammers2015}.
This localization becomes more pronounced with increasing $\Delta n_p$ (panel f), with activity appearing to emanate exclusively
from a source at the upper end of the domain (the exact location of the
source varies depending on the realization of the $n_p$
distribution). 
These results 
demonstrate that
unidirectional propagation of excitation waves 
can arise in a system with a gradient in the density of 
coupling between non-oscillatory cells.
This can be explicitly seen from Fig.~\ref{fig_kymo_annealed}~(a) which
displays the space-time evolution of membrane potential
along a longitudinal section of the simulation domain (indicated by the
broken lines in Fig.~\ref{fig_annealed}, panels d-f). 
When 
$\Delta n_p = 0$,
activity is seen to be initiated by multiple sources located at
different parts of the domain (left panel). For finite $\Delta n_p$,
a single source located at the upper end drives activity across the
domain, with waves originating from this site propagating through the entire
medium (center and right panels). 
As already mentioned above, the number of organizing centers decreases with increasing slope $\Delta n_p$ of the passive cell density gradient and/or the inter-cellular
coupling strength $D (=C_r)$. We quantitatively establish this in terms
of the variation in the
number of phase singularities $n_s$ corresponding to the spiral tips
as a function of these parameters 
(Fig.~\ref{fig_kymo_annealed},~b)~\cite{note1}. 
Taken together, these results demonstrate that the existence of a 
centralized pacemaker region near the fundus is not necessary to
explain the coordination of uterine activity.

The phase velocity $v$ of the propagating excitation front is seen to be a decreasing function of the distance
$d$ from the organizing center (Fig.~\ref{fig_kymo_annealed},~c)~\cite{note3}.
It reflects the decrease in the intrinsic frequencies of the oscillators
arising from the reduction in $n_p$ along the longitudinal
axis.
As can be seen from the figure, $v$
drops with $d$ more sharply as $\Delta n_p$ is increased. 
We note that this observation suggests a testable prediction of
the mechanism proposed here for the coordination of uterine activity.
Specifically, a spatial gradient of cellular coupling being responsible for
the observed phenomena cannot be ruled out unless the phase velocity for propagating activity on the uterus is shown to be independent of the distance from a putative pacemaker region. Moreover, by measuring the dependence of the phase velocity on this distance, 
it may be possible to infer the steepness of this gradient empirically.

Further increase of $\Delta n_p$ beyond the range considered so far
eventually results in partial cessation of activity in the medium.
Fig.~\ref{fig_kymo_annealed}~(d) shows that when the
slope $\Delta n_p \geq 2 \times 10^{-3}$, the fraction of oscillating cells
in the lattice begins to reduce from $1$ as the number of passive cells 
at the lower end of the density gradient become insufficient for supporting
spontaneous activity. We note that in the limit of extremely high $\Delta n_p$
(corresponding to the passive cell density varying as a step function),
the system reduces to one having all oscillating cells
confined in the upper segment, with no activity propagating to the
lower segment of the domain. While the oscillating region at the top 
has the potential to effectively function as a pacemaker, its inability
to activate the rest of the medium rules this out as a plausible 
mechanism operating in the uterus~\cite{note2}.

Recent studies have suggested
that for smaller mammals, such as guinea pigs, activity can
arise from anywhere within the myometrium~\citep{Lammers2015}. 
However, it has been pointed out that successful vaginal delivery
in humans is usually associated 
with fundal dominance~\citep{Alvarez1954,Buhimschi2009,Euliano2009}.
Our results suggest that coherence can be achieved even in the absence
of a spatially localized coordinating center in animals whose uteri 
have linear dimensions 
smaller than the wavelength of the propagating activity.
Thus, a spatial symmetry-breaking gradient may be
crucial only for animals with extended uteri, where co-existing wave
sources need to be coordinated in
order to achieve coherent contraction.
Indeed, arrested labor may result if the source of activity
is located far below the fundus in the human uterus~\citep{Euliano2009},
and in some cases, the presence of multiple
sources may also cause
uterine fibrillation~\citep{Alvarez1954}. 
Furthermore, we note that the location of the coordination centers in
different realizations can vary because of the stochastic nature of
the $n_p$ distribution along the gradient. This suggests
a possible explanation for the continuing uncertainty regarding the position 
of the foci of uterine activity in experimental studies~\cite{Lammers2013, Rabotti2015}.

\begin{acknowledgments}
This research was supported in part by IFCAM. PS has been supported
by NNMCB (a program of the Science \& Engineering Research Board, 
Government of India) and the IMSc Complex Systems 
Project (12th Plan) funded by the Department of Atomic Energy, Government of
India.
SNM has been supported by the IMSc Complex Systems Project (12th Plan), and 
the Center of Excellence in Complex Systems and Data 
Science, both funded by the Department of Atomic Energy, Government of 
India.
The simulations and computations required for this work were supported by
the Institute of Mathematical Sciences
High Performance Computing facility (hpc.imsc.res.in) [Nandadevi
and Satpura clusters]. 
\end{acknowledgments}


\begin{thebibliography}{100}
\bibitem{note1}
Phase singularities are identified as the points of intersection of
the isoclines of the $V_e$ field at two different times separated by
a small interval.
\bibitem{note3}
To prevent the stochastic
fluctuations in the passive cell density gradient from obscuring the
relation between phase velocity and the distance of the wave
from the organizing center, we have obtained it along a one-dimensional
section parallel to the gradient with the $n_p$ at site $l$ equal to
$\lambda(l) (= [l-(L/2)] \Delta n_p + n_p^{mid})$.
\end{thebibliography}
\end{document}